\documentclass[12pt]{article}

\usepackage{kotex}
\usepackage{graphicx}
\usepackage{array}
\usepackage{tabularx}
\usepackage{amsmath,amssymb,bbold}
\usepackage{url}
\usepackage{rotating}
\usepackage{color}
\usepackage{hhline}
\usepackage{hyperref}
\usepackage{slashed}
\usepackage{makecell}
\usepackage{empheq}
\usepackage{physics}
\usepackage{comment}
\usepackage{authblk}
\usepackage{xcolor}
\usepackage[dvips=false,pdftex=false,vtex=false]{geometry}
\usepackage{changepage}
\usepackage{simpler-wick}
\usepackage{makecell}
\usepackage{multirow}
\usepackage{cleveref}
\usepackage{cite}
\usepackage{setspace}
\begin{document}

\newcommand{\s}{\\ \vspace*{-2mm}}

\begin{flushright}
KIAS-Q25002
\end{flushright}

\vskip 1.cm

\begin{center}
{\Large \bf No gauge cancellation at high energy
\\[5pt]
in the five-vector $R_\xi$ gauge}\\[20pt]
{Jaehoon Jeong\footnote{jeong229@kias.re.kr, jaehoonjeong229@gmail.com}} \\[15pt]
{\it  Quantum Universe Center, KIAS, Seoul 02455, Korea}
\end{center}

\vskip 0.5cm

%
%

\begin{abstract}
We propose a novel $R_\xi$ gauge in the five-vector (5V) framework within the Abelian Higgs model. In the Cartesian basis of the complex Higgs field, the 5V $R_\xi$ gauge ensures non-divergent tree-level amplitudes for each Feynman diagram in the high-energy limit. This framework pinpoints the origin of high-energy divergences in tree-level amplitudes for each diagram, providing a criterion for quantifying the degree of divergences from other gauges. The 5V description necessitates treating the Goldstone field as the fifth gauge-field component, offering deeper insight into the dynamics of massive gauge bosons, particularly its longitudinal mode. The impact of this framework is demonstrated by rigorously comparing tree-level amplitudes from the 5V $R_\xi$ gauge with those from the conventional 4V $R_\xi$ gauge and the Feynman diagram gauge, the latter of which exhibits no gauge cancellation, similar to the 5V $R_\xi$ gauge.
\end{abstract}

\section{Introduction}

In gauge theories, gauge cancellation~\cite{Lee:1977eg} plays a critical role in ensuring unitarity at high energies by effectively nullifying divergences in amplitudes arising from the longitudinal polarizations of gauge bosons.
While various gauge-fixing schemes exist, addressing high-energy divergences and gauge cancellation remains a critical challenge.
To tackle this issue, a novel gauge-fixing method in the 5V framework, initially known as the Equivalent gauge~\cite{Wulzer:2013mza,Cuomo:2019siu}, was developed through covariant quantization. Subsequently, the authors of Refs.~\cite{Hagiwara:2020tbx,Chen:2022gxv,Chen:2022xlg,Hagiwara:2024xdh,Hagiwara:2024ron} provided an alternative derivation of this gauge by obtaining its propagator directly from the Green’s function of the equation of motion (EOM), which led to its rebranding as the Feynman-diagram (FD) gauge. Their work primarily focused on addressing numerical challenges in scattering simulations, particularly in handling high-energy divergences in scattering amplitudes, with various practical applications. This approach ensures that tree-level amplitudes for each Feynman diagram remain finite even in the high-energy limit, representing a crucial advancement in stabilizing simulations and improving numerical accuracy in scattering calculations.
In the FD gauge, the longitudinal polarization vectors are defined using a light-like gauge vector, whose momentum-dependent components are normalized to be dimensionless. However, despite its utility, the exotic structure of the gauge vector complicates the formulation of a simple gauge-fixing term in the Lagrangian, making it challenging to systematically analyze its physical implications in a simple way.
\s

The Becchi-Rouet-Stora-Tyutin (BRST) symmetry~\cite{Becchi:1975nq, Tyutin:1975qk}, which is essential for the correct quantization of gauge fields, ensures that gauge redundancies are consistently accounted for in the path integral formalism, effectively eliminating nonphysical degrees of freedom. In particular, it establishes a connection between the nonphysical polarization modes of the gauge boson and the contribution from the Nambu-Goldstone boson~\cite{Nambu:1960tm, Goldstone:1961eq, Valencia:1990jp}. Regardless of the gauge-fixing method, a current $T_{\mu}(A)$ coupled to the polarization vectors of a massive gauge boson $A$ with mass $m$ can be expressed as~\cite{Ruegg:2003ps}
\begin{align}
p^{\mu} T_{\mu}(A)=\mp i m T (\pi),
\end{align}
in terms of the amplitude $T(\pi)$ of the Goldstone boson $\pi$, where the signs indicate the incoming ($-$) or outgoing ($+$) gauge boson $A$, with momentum $p$. This formula suggests that currents associated with massive gauge bosons exhibit modified conservation properties in the four-dimensional framework. Given that the currents associated with massless gauge bosons play a crucial role in ensuring the full absence of unphysical momentum dependence, it would be beneficial to preserve current conservation even for massive gauge bosons in order to identify unphysical contributions by adopting the 5V framework.
\s

In this work, we present a novel $R_\xi$ gauge-fixing approach in the 5V framework within the Abelian Higgs model~\cite{Higgs:1964pj,Englert:1964et}, where the gauge-fixing term is a straightforward extension of the 4V $R_\xi$ gauge to the 5V framework. By regarding the Goldstone field as the fifth gauge-field component, we derive the three physical polarization vectors, all of which are transverse to the five-momentum. Unlike the FD gauge, where the Goldstone component remains a constant imaginary value regardless of energy, the polarization-vector structure in the 5V $R_\xi$ gauge exhibits a striking energy dependence: in the rest frame, the fifth component completely vanishes, whereas in the high-energy or massless limit, all four-vector components disappear, leaving only the fifth component. Thus, the structure of the third polarization vector clearly demonstrates that it is not an intrinsic property of the gauge boson, exhibiting different spin-angular momentum values transitioning from spin-1 in the rest frame to spin-0 in the high-energy or massless limit.
\s

It is shown that the 5V $R_\xi$ gauge yields tree-level amplitudes equivalent to those in the 4V $R_\xi$ and FD gauges in the polar basis for the complex Higgs field, where the amplitudes for each diagram diverge at high energy regardless of the gauge choice. This framework fully identifies the origin of these polar-basis divergences by examining the equivalence between the high-energy and vanishing vacuum expectation value (VEV) limits, ensuring no gauge cancellation in the Cartesian basis for the complex Higgs field. However, the 4V $R_\xi$ gauge leads to divergent high-energy amplitudes, even in the Cartesian basis. The simple analytic structure of the 5V description allows us to easily trace these divergences and identify the origin of the absence of gauge cancellation in the FD gauge, emphasizing the practical advantages of the 5V framework as an effective tool for identifying high-energy divergences from other gauges.
\s

After the introduction, Section~\ref{sec:r_xi_gauge} derives the 5V $R_\xi$ gauge-fixing term in the Lagrangian of the Abelian Higgs model with the complex Higgs field in the polar basis, obtained by imposing BRST invariance on the Lagrangian including the gauge-fixing term. Subsequently, the three physical polarization vectors, which are transverse to the five-momentum, are determined by analyzing the state operators and identifying the unphysical ones that do not belong to the cohomology. Afterward, the physical implications of the third transverse mode, corresponding to the longitudinal mode in the 4V description, are examined in detail. Section~\ref{sec:equivalence_5V_4V} examines the tree-level amplitudes for the processes $A_LA_L \rightarrow A_LA_L$ and $A_LA_L \rightarrow hh$, involving massive gauge bosons $A_L$ with longitudinal polarization and Higgs bosons $h$. It is shown that the tree-level amplitudes for each Feynman diagram, derived from the 5V $R_\xi$ gauge, are equivalent to those from the 4V $R_\xi$ and FD gauges in the polar basis. Section~\ref{sec:identification_high-energy} presents the high-energy divergences in the amplitudes through explicit computations and identifies their origin from the equivalence between the high-energy and vanishing VEV limits, ensuring no gauge cancellation in the Cartesian basis for the complex Higgs field. After discussing the rationale for the high-energy divergences in the 4V $R_\xi$ gauge and the origin of no gauge cancellation in the FD gauge,  the impact of the 5V $R_\xi$ gauge is explored through a rigorous comparison of the amplitudes from the 5V $R_\xi$, 4V $R_\xi$, and FD gauges. Section~\ref{sec:conclusion} summarizes the key findings and examines the physical implications of the 5V framework. The ingredients for the amplitude calculations are provided in Appendix~\ref{appendix:kinematics}.

\section{5V $R_\xi$ gauge in the Abelian Higgs model}
\label{sec:r_xi_gauge}

In this section, we establish the 5V $R_\xi$ gauge in the Abelian Higgs model which is the simplest example among the models including massive gauge bosons of which the masses are generated via the Higgs mechanism. Deriving the physical polarization vectors representing the three transverse polarization modes of the massive gauge boson in the 5V description, we trace the identity of the longitudinal polarization mode in the context of the 5V framework. Let us consider first the Lagrangian in the Abelian Higgs model with the complex Higgs boson in the polar basis after spontaneous symmetry breaking (SSB) as
\begin{align}
\mathcal{L}(x)
=&-\frac{1}{4} (F^{\mu\nu})^2
+|D^\mu \Phi|^2
-\frac{\lambda}{4}\bigg(|\Phi|^2 - \frac{2\mu^2}{\lambda}\bigg)^2,
\label{eq:Lagrangian}
\end{align}
where $F^{\mu\nu}=\partial^\mu A^\nu
-\partial^\nu A^\mu$ is the field strength tensor of the gauge field $A$, $D^\mu=\partial^\mu-ieA^\mu$ is the covariant derivative, and $\Phi$ is the complex Higgs field. Setting the polar basis $\Phi=\frac{1}{\sqrt{2}}(v+h)e^{i\pi/v}$ with the Goldstone field $\pi$, we can expand the Lagrangian as
\begin{align}
\mathcal{L}(x)
=&-\frac{1}{4} F^{\mu\nu}F_{\mu\nu}
+\frac{1}{2}m^2\bigg(A^{\mu}-\frac{1}{m}\partial^\mu \pi \bigg)^2
+\frac{1}{2}(\partial^{\mu} h)^2 - \frac{1}{2} m_h^2 h^2
\nonumber
\\
&+\bigg(\frac{\lambda^{1/2} m^2}{\sqrt{2} m_h}h
+\frac{\lambda m^2}{4m_h^2}h^2\bigg) 
\bigg(A^{\mu}-\frac{1}{m}\partial^\mu \pi \bigg)^2 
-\frac{\lambda^{1/2}m_h}{2\sqrt{2}}  h^3 - \frac{\lambda}{16} h^4,
\label{eq:Lagrangian_expanded}
\end{align}
in terms of the VEV $v=2\mu/\sqrt{\lambda}$, where $m=ev$ and $m_h=\lambda^{1/2} v/\sqrt{2}$ are the masses of the gauge and Higgs bosons.  This Lagrangian is invariant under the gauge transformation $A^\mu \rightarrow A^\mu +\partial^\mu \Lambda$ and $\pi \rightarrow \pi +m\Lambda$. If we regard the Goldstone field $\pi$ as the fifth gauge-field component $A^4$, we can obtain
\begin{align}
\mathcal{L}(x)
=&-\frac{1}{4} F^{MN}F_{MN}+\frac{1}{2} (\partial^{\mu} h)^2 - \frac{1}{2} m_h^2 h^2
\nonumber
\\
&+\bigg(\frac{\lambda^{1/2} m^2}{\sqrt{2} m_h}h
+\frac{\lambda m^2}{4m_h^2}h^2\bigg) 
\bigg(A^{\mu}-\frac{1}{m}\partial^\mu A^4 \bigg)^2 
-\frac{\lambda^{1/2}m_h}{2\sqrt{2}}  h^3 - \frac{\lambda}{16} h^4,
\label{eq:5V_Lagrangian}
\end{align}
in terms of the 5V field strength tensor $F^{MN}=\partial^M A^N-\partial^N A^M$ with the 5V indices $M,N=0,1,2,3,4$ including the 5V gauge field $A^M$. Here, the five-dimensional Minkowski metric tensor and mass-included 5V derivative involved by $F^{MN}$ are given by
\begin{align}
g_{MN}&=\mbox{diag}(1,-1,-1,-1,-1),
\\
\partial^M&=(\partial^\mu,m).
\end{align}
Note that the Lagrangian in Eq.~\eqref{eq:5V_Lagrangian} expressed in the 5V description is invariant under the gauge transformation $A^M \rightarrow A^M+\partial^M \Lambda$. 
\s 

To ensure the invariance of the $S$-matrix under the addition of the gauge-fixing term, the Lagrangian, including both the gauge-fixing and ghost terms, must remain invariant under the BRST transformation. In the 5V framework, the BRST transformations of the fields can be expressed in terms of the BRST operator $\mathbf{s}$ as
\begin{align}
\mbox{\bf s} A^M&=[iQ_B,A^M]=\partial^M c,
\label{eq:brst_polar1}
\\
\mbox{\bf s}\, h&=[iQ_B,h]=0,
\label{eq:brst_polar2}
\\
\mbox{\bf s} B&=[iQ_B,B]=0,
\label{eq:brst_polar3}
\\
\mbox{\bf s}\, \bar{c}&=\{iQ_B,\bar{c}\}=B,
\label{eq:brst_polar4}
\\
\mbox{\bf s}\, c&=\{iQ_B,c\}=0,
\label{eq:brst_polar5}
\end{align}
in terms of the BRST charge $Q_B$, where $B$ is the commuting Lautrup-Nakanishi auxiliary field~\cite{Lautrup:1967zz,Nakanishi:1966zz}, and $c$ and $\bar{c}$ are the anticommuting ghost and anti-ghost fields, respectively. It is straightforward to check the BRST invariance of each term in the Lagrangian~\eqref{eq:5V_Lagrangian}. With an arbitrary local functional $\mathcal{G}$ depending on the fields, we can construct the BRST-invariant term which is the combination of the gauge-fixing and ghost-field terms:
\begin{align}
\mathcal{L}_{\scriptsize\mbox{gf+gh}}
=\mbox{\bf s}\bigg[\bar{c}\bigg(\mathcal{G}(A,h,B)+\frac{\xi}{2}B\bigg) \bigg],
\end{align}
in terms of the gauge-fixing parameter $\xi$, satisfying $\mbox{\bf s} \mathcal{L}_{\scriptsize\mbox{gf+gh}}(x)=0$. The auxiliary field $B$ can be eliminated by using its own EOM as
\begin{align}
\frac{\partial \mathcal{L}_{\scriptsize\mbox{gf+gh}}}{\partial B}
=0 \;\; \rightarrow \;\; B=-\frac{\mathcal{G}}{\xi},
\end{align}
so that
\begin{align}
\mathcal{L}_{\scriptsize\mbox{gf+gh}}
=-\frac{\mathcal{G}^2}{2\xi}-\bar{c}\big(\mbox{\bf s} \,\mathcal{G}\big).
\end{align}
The 5V $R_\xi$ gauge-fixing term can be established by defining the functional $\mathcal{G}$ as 
\begin{align}
\mathcal{G}(A)=\bar{\partial}_{M} A^M,
\end{align}
where $\bar{\partial}^M=(\partial^\mu,-m)=\partial_M$, referred to as the conjugated 5V derivative. This results in the combination to be given by
\begin{align}
\mathcal{L}_{\scriptsize\mbox{gf+gh}}
=-\frac{(\bar{\partial}_{M} A^M)^2}{2\xi}-\bar{c}(\bar{\partial}^M\partial_M)c.
\label{eq:5V_gauge-fixing}
\end{align}
It follows that the primary difference between the 5V $R_\xi$ gauge and the conventional 4V $R_\xi$ gauge arises from the definition of the functional $\mathcal{G}$, where the fifth component does not involve the gauge-fixing parameter $\xi$, unlike in the 4V $R_\xi$ gauge.
\s

To construct the propagator of the 5V gauge field, let us employ the kinetic term of the gauge boson in the Lagrangin~\eqref{eq:5V_Lagrangian} with the gauge fixing-term in Eq.~\eqref{eq:5V_gauge-fixing}:
\begin{align}
\mathcal{L}_{\scriptsize\mbox{free}} = -\frac{1}{4} F^{MN}F_{MN} -\frac{(\bar{\partial}_{M} A^M)^2}{2\xi}.
\label{eq:free_plus_gauge-fixing_L}
\end{align}
Employing this expression, we can find the free-theory EOM as
\begin{align}
\begin{pmatrix}
(\square +m^2)g^\mu_{\;\,\nu}-\Big(1-\frac{1}{\xi}\Big)
\partial^\mu \partial_{\nu}
& -m\Big(1-\frac{1}{\xi}\Big) \partial^\mu
\\
-m\Big(1-\frac{1}{\xi}\Big) \partial_\nu& \square +\frac{m^2}{\xi}
\end{pmatrix}
\begin{pmatrix}
A^\mu(x) \\ A^4(x)
\end{pmatrix}
=0.
\end{align}
Fourier transforming to momentum space leads to
\begin{align}
O^M_{\;\;\, N}\tilde{A}^N(P)=0
\quad
\mbox{with}
\quad
O^M_{\;\;\, N}=-(P^{R} P_{R}^*)\, g^M_{\;\;\,N}
+\bigg(1-\frac{1}{\xi}\bigg)P^{M}P_{N}^*,
\end{align}
in terms of the Fourier-transformed gauge field $\tilde{A}^M (P)$, where the five-momentum $P$ and its scalar product are cast by
\begin{align}
P^M&=(p^0,\vec p, im)\;\; \mbox{and} \;\; P^{R} P_{R}^*=(p^0)^2-\vec p^{\,2}-m^2.
\end{align}
Here, the scalar product is defined with a five-vector and a conjugated five-vector. It is straightforward to find the 5-by-5 matrix $ K^M_{\;\;\, N} $ that satisfies the identity:
\begin{align}
K^M_{\;\;\,N}O^N_{\;\;\,R}A^R = P^2 \,g^{M}_{\;\;\,R}A^R,
\end{align}
with the abbreviation $ P^2 = P^M P_{M}^* $ where the matrix $ K^M_{\;\;\, N} $ is expressed as:
\begin{align}
K^M_{\;\;\,N} = -g^{M}_{\;\;\,N} + (1-\xi)\frac{P^M P^*_{N}}{P^2} .
\end{align}
Then, the 5V propagator, which serves as the Green's function of the EOM, can be succinctly represented by
\begin{align}
i\Pi_{MN} = \frac{iK_{MN}}{P^2}=\frac{\displaystyle i 
\bigg(-g_{MN} + (1-\xi)\frac{P_M P^*_{N}}{P^2}\bigg)}{P^2},
\label{eq:5V_propagator}
\end{align}
indicating the presence of a light-like pole at $ P^2 = (p^{0})^2 - \vec{p}^{\,2} - m^2 = 0 $ in the 5V description.
\s

To identify the physical polarization of a gauge boson in the 5V description, let us take the quantization of the gauge and ghost fields  as~\cite{Srednicki_2007}
\begin{align}
A^M&=\int \frac{d^3 \vec p}{(2\pi)^3\sqrt{2E}}
\sum_{\tiny
\begin{minipage}{1cm}
\mbox{{\scriptsize $\lambda$}\,$=\pm 1,0,$ }
\\ 
\mbox{\qquad $>,<$}
\end{minipage}}
\Big[\epsilon^M (p,\lambda) a(p,\lambda)e^{-i p\cdot x}
+ \epsilon^{M*} (p,\lambda) a^\dagger(p,\lambda)e^{i p\cdot x} \Big],
\label{eq:vector_quantization}
\\
c&=\int \frac{d^3 \vec p}{(2\pi)^3\sqrt{2E}}
\Big[C(p)e^{-ip\cdot x}+C^\dagger(p) e^{ip\cdot x} \Big],
\label{eq:c_quantization}
\\
\bar{c}&=\int \frac{d^3 \vec p}{(2\pi)^3\sqrt{2E}}
\Big[\bar{C}(p)e^{-ip\cdot x}+\bar{C}^\dagger(p) e^{ip\cdot x} \Big],
\label{eq:c_bar_quantization}
\end{align}
with the on-shell energy $E=\sqrt{\vec p^{\;2}+m^2}$. Recall that the Lagrangian in Eq.~\eqref{eq:5V_Lagrangian} is gauge-invariant under the transformation $A^M \rightarrow A^M + \partial^M \Lambda$, which implies that the polarization in the direction of the five-momentum is unphysical. This motivates us to choose the polarization vectors in Eq.~\eqref{eq:vector_quantization} based on the five-momentum. Our choice consists of three space-like and two light-like polarization vectors, which together span the five-dimensional momentum space:
\begin{align}
\epsilon^M(p,\pm 1)
&=\frac{1}{\sqrt{2}}(0,\mp \hat{\theta}-i\hat{\phi},0),
\label{eq:conventional_transverse}
\\
\epsilon^M(p,0)
&=\frac{1}{E}(0,m\hat{p},-i |\vec p\,|),
\label{eq:third_transverse}
\\
\epsilon^M(p,>)
&=\frac{1}{\sqrt{2}E}(E,\vec p, im),
\\
\epsilon^M(p,<)
&=\frac{1}{\sqrt{2}E}(E,-\vec p, -im),
\end{align}
where the 3V spatial unit vectors are given by
\begin{align}
\hat{p}&=(\sin\theta \cos\phi,\sin\theta \sin\phi,\cos\theta),
\\
\hat{\theta}&=(\cos\theta \cos\phi,\cos\theta \sin\phi,-\sin\theta),
\\
\hat{\phi}&=(-\sin\phi,\cos\phi,0),
\end{align}
in terms of the polar $\theta$ and azimuthal $\phi$ angles. Here, the space-like polarization vectors, $\epsilon(\pm)$ and $\epsilon(0)$, satisfy the following orthonormality,
\begin{align}
\epsilon^M(p,\sigma)\epsilon^*_M(p,\sigma')=-\delta_{\sigma,\sigma'},
\end{align}
which are the eigenstates over the helicity operator, yielding the eigenvalues, $\pm 1$ and $0$, respectively. It indicates their correspondence to the conventional transverse and longitudinal polarization vectors in the 4V description. Here, $\epsilon^M(\pm 1)$ is represented by adopting the Wick convention~\cite{Wick:1962zz}. The light-like vectors, $\epsilon(>)$ and $\epsilon(<)$, are spatially parallel and opposite to the five-momentum $P$, respectively. Employing the BRST transformations of the fields in Eqs.~\eqref{eq:brst_polar1} to \eqref{eq:brst_polar5}, we can find straightforwardly the following relations for the operators defined in Eqs.~\eqref{eq:vector_quantization} to \eqref{eq:c_bar_quantization} as 
\begin{align}
[Q_B, a^\dagger_{\lambda}(p)]&=
\sqrt{2}E\, \delta_{\lambda,>} C^\dagger(p),
\label{eq:brst_trans_operator_form1}
\\
\{Q_B, \bar{C}^\dagger(p)\}&= -\xi^{-1}\sqrt{2}E\, a^\dagger_< (p),
\label{eq:brst_trans_operator_form2}
\end{align}
which are valid even in the massless limit ($m \to 0$). These relations allow us to eliminate the unphysical polarization modes corresponding to the light-like vectors with indices $>$ and $<$ in the context of cohomology; any physical state must be in the cohomology~\cite{Becchi:1975nq,Tyutin:1975qk}. A physical state $|\Psi \rangle$ defined as a normalized and BRST-closed state, satisfying $\langle \Psi |\Psi \rangle = 1$ and $Q_B |\Psi \rangle = 0$. The normalization condition implies that a physical state $|\Psi\rangle$ cannot be written in the BRST-exact form $Q_B |\Psi'\rangle$ for any state $|\Psi'\rangle$, since such a state has vanishing norm as a consequence of the nilpotency $Q_B^2 = 0$ of the Hermitian BRST operator $Q_B$. According to Eq.~\eqref{eq:brst_trans_operator_form1}, the state $a^\dagger_>(p) |\Psi \rangle$ is not BRST-closed, since it is mapped by $Q_B$ to a nonzero state: $Q_B a^\dagger_>(p) |\Psi \rangle = \sqrt{2}E\, C^\dagger(p) |\Psi \rangle$. On the other hand, Eq.~\eqref{eq:brst_trans_operator_form2} shows that $a^\dagger_<(p) |\Psi \rangle$ is proportional to $Q_B \bar{C}(p) |\Psi \rangle$, implying that it is BRST-exact and thus does not belong to the cohomology.
\s

Then, we can conclude that in the 5V description, a gauge boson possesses three physical polarization modes with helicities $\pm 1$ and $0$, of which the polarization vectors in Eqs.~\eqref{eq:conventional_transverse} and \eqref{eq:third_transverse} are transverse to the five-momentum. Recall our treatment of the Goldstone field as the fifth gauge-field component. The third transverse mode corresponding to the longitudinal mode reflects the dominance of the Goldstone boson contribution to amplitudes in the high-energy limit, known as the Goldstone-boson equivalence theorem~\cite{Goldstone:1961eq}.
\s

In quantum mechanics, an angular momentum state is defined to possess the same angular momentum, regardless of its projected value along an axis. However, the third transverse polarization vector exhibits different spin-angular momentum values: 1 in the rest frame ($E = m$) and 0 in the high-energy ($E \rightarrow \infty$) or massless limit ($m \rightarrow 0$). This strongly suggests that the third transverse mode of a gauge boson does not involve an intrinsic property that remains invariant under Lorentz transformations. In other words, its property is dependent on the five-momentum, resulting in a transition between different types of physical states under Lorentz transformations.

\section{Connecting the 5V $R_\xi$ gauge with other gauges}
\label{sec:equivalence_5V_4V}

We demonstrate that the tree-level amplitudes from the 5V $R_\xi$ gauge are equivalent to those from the 4V $R_\xi$ and FD gauges in the polar basis. By employing the Lagrangian~\eqref{eq:5V_Lagrangian}, we can construct the Feynman rules for the interaction vertices as follows:
\begin{align}
\Gamma^{[AAh]}_{M_1 M_2}(P_1,P_2)&= 
\frac{\sqrt{2\lambda} m^2}{m_h P_{1,4}P_{2,4}}
\begin{pmatrix}
\displaystyle P_{1,4}P_{2,4}\, g_{\mu_1\mu_2}& \displaystyle  P_{1,4} P_{2\mu_1}
\\[15pt]
\displaystyle   P_{2,4} P_{1\mu_2}& \displaystyle P_1^\mu P_{2\mu}
\end{pmatrix}, 
\label{eq:polar_vertex1}
\\
\Gamma^{[AAhh]}_{M_1 M_2}(P_1,P_2)&=\frac{\lambda m^2}{m_h^2P_{1,4}P_{2,4}}
\begin{pmatrix}
\displaystyle P_{1,4}P_{2,4}\, g_{\mu_1\mu_2}& \displaystyle  P_{1,4} P_{2\mu_1}
\\[15pt]
\displaystyle   P_{2,4} P_{1\mu_2}& \displaystyle P_1^\mu P_{2\mu}
\end{pmatrix}, 
\label{eq:polar_vertex2}
\\
\Gamma^{[hhh]}&=-\frac{3\lambda^{1/2}m_h}{\sqrt{2}} ,
\label{eq:polar_vertex3}
\\
\Gamma^{[hhhh]}&=-\frac{3\lambda }{2},
\label{eq:polar_vertex4}
\end{align}
for the interactions, $AAh$, $AAhh$, $hhh$, and $hhhh$, of the gauge and Higgs bosons, $A$ and $h$, where the lower-index fifth components are defined as $P_{1,4}=g_{4 M} P^M_1$ and $P_{2,4}=g_{4 M} P^M_2$, from the five-momenta, $P_1$ and $P_2$, respectively. Each of the simplified matrix expressions in Eqs.~\eqref{eq:polar_vertex1} and~\eqref{eq:polar_vertex2} represents a 4 by 4 matrix in the top left, four-component column and row vectors in the top right and bottom left, and a scalar element in the bottom right. The interaction vertices are divergence-free:
\begin{align}
&P_{1}^{M_1} \Gamma_{M_1M_2}(P_1,P_2)=P_{2}^{M_2} \Gamma_{M_1M_2}(P_1,P_2)=0 \!\!\!\!\!\!\!\! && \!\!\!\!\!\!\!\!
\mbox{for incoming $A$'s},
\label{eq:divergence_free_incoming}
\\
&P_{1}^{M_1*} \Gamma_{M_1M_2}(-P_1^*,-P^*_2)=P_{2}^{M_2*} \Gamma_{M_1M_2}(-P_1^*,-P^*_2)=0 \!\!\!\!\!\!\!\! && \!\!\!\!\!\!\!\!
\mbox{for outgoing $A$'s},
\label{eq:divergence_free_outgoing}
\end{align}
reflecting the 5V gauge symmetry of the Lagrangian in Eq.~\eqref{eq:5V_Lagrangian}.
These ensure the Ward identity~\cite{Ward:1950xp,Takahashi:1957xn} in the 5V description, implying the conservation of any current coupled to the 5V gauge field. The outgoing momenta can be derived straightforwardly via the complex conjugation.
\s

\begin{figure}[ht!]
\centering
\includegraphics[scale=1.4]{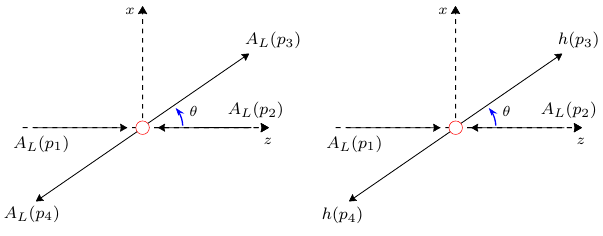}
\caption{Kinematics configurations for the 2-to-2 scattering processes $A_LA_L\rightarrow A_LA_L$ (Left) and $A_LA_L\rightarrow hh$ (Right) of the longitudinally polarized massive gauge bosons $A_L$ and the Higgs bosons $h$. $p_{1,2,3,4}$ are four momenta for each particle, and $\theta$ is the polar angle defined with respect to the $z$ axis.}
\label{fig:kinematics}
\end{figure}

\begin{figure}[ht!]
\centering
\includegraphics[scale=1.0]{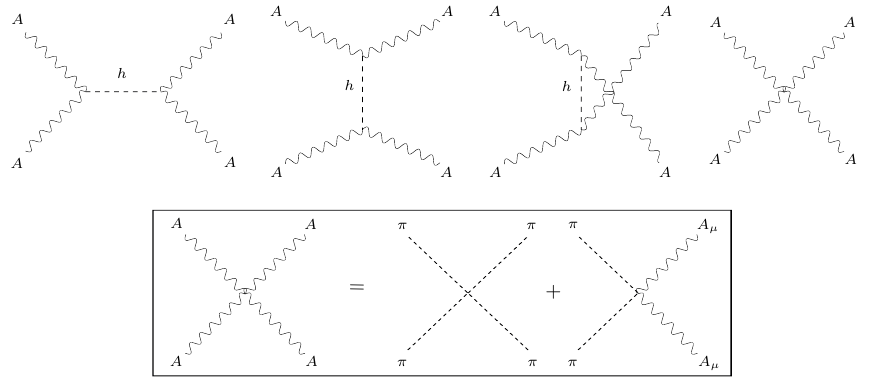}
\caption{Feynman diagrams for the tree-level process $AA\rightarrow AA$. For clarity, we include the contact interaction diagram with the Goldstone $\pi$ and gauge $A_\mu$ bosons in the conventional description, appeaing only in the Cartesian basis.}
\label{fig:AAAA_diagrams}
\end{figure}

The demonstration of the straightforward transformation begins by evaluating the amplitudes for the $A_L A_L \rightarrow A_L A_L$ scattering process in the center of mass (c.m.) frame (see the kinematic configuration in the left panel of Fig.~\ref{fig:kinematics}), specifically for four longitudinally polarized massive gauge bosons.
This process involves three Feynman diagrams: the $s$-, $t$-, and $u$-channel processes, including the Higgs boson exchange as described in Fig.~\ref{fig:AAAA_diagrams}. 
Despite no actual computation of the amplitudes, we can easily transform the amplitudes from the 5V $R_\xi$ gauge to those from different gauges by re-expressing the third transverse polarization vector $\epsilon^M(p,0)$ in Eq.~\eqref{eq:third_transverse} with the five-momentum $P$ as
\begin{align}
\epsilon^{M}(p,0)=\epsilon^{M}_0(p,0)-\frac{|\vec p\,|}{E}\frac{P^M}{m} \;\;\mbox{with}\;\; 
\epsilon^M_{0}(p,0)=\frac{1}{m}(|\vec{p}\,|,E\hat{p},0),
\label{eq:5V_rep_of_longitudinal_vector}
\end{align}
in terms of the on-shell energy $E = \sqrt{\vec{p}^{\,2} + m^2}$. This ensures that the tree-level amplitudes for the process $A_L A_L \rightarrow A_L A_L$, computed with $\epsilon(0)$'s, are equivalent to those computed with $\epsilon_0(0)$'s by the divergence-free vertices in Eqs.~\eqref{eq:divergence_free_incoming} and \eqref{eq:divergence_free_outgoing}. Using $\epsilon_0(p,0)$, and with the definitions $\epsilon^\mu(p, \pm 1) = \epsilon^\mu_0(p, \pm 1)$, we can derive the identities connecting the amplitudes from the 5V and 4V $R_\xi$ gauges. For instance, in the $s$-channel Higgs exchange, the scalar current of the massive gauge boson, coupled to the Higgs boson, satisfies
\begin{align}
\epsilon^{M_{1}}(p_{1},\lambda_1)
\Gamma^{[AAh]}_{M_1M_2}(P_1,P_2)
\epsilon^{M_{2}}(p_{2},\lambda_2) &= 
\epsilon_{0}^{\mu_{1}}(p_{1},\lambda_1)
\Gamma^{[AAh]}_{\mu_1\mu_2}(P_1,P_2)
\epsilon_{0}^{\mu_{2}}(p_{2},\lambda_1),
\\
\epsilon^{M_{3}*}(p_{3},\lambda_3)
\Gamma^{[AAh]}_{M_3M_4}(-P_3^*,-P_4^*)
\epsilon^{M_{4}*}(p_{4},\lambda_4) &= 
\epsilon_{0}^{\mu_3*}(p_{3},\lambda_3)
\Gamma^{[AAh]}_{\mu_3\mu_4}(-P_3^*,-P_4^*)
\epsilon_{0}^{\mu_4*}(p_{4},\lambda_4),
\end{align}
in terms of arbitrary helicities, $\lambda_{1,2,3,4} = \pm 1$ or $0$, where $P_{1,2}^M = (p_{1,2}^\mu, i m)$ and $P^{M*}_{3,4} = (p_{3,4}^\mu, -im)$ are the five-momenta of four massive gauge bosons. These result in identical tree-level amplitudes in both the 5V and 4V descriptions. The explicit expressions of the $A_L A_L \rightarrow A_L A_L$ amplitudes for each diagram and their sum (see the computation ingredients in Appendix.~\ref{appendix:kinematics}) can be found in the second column of Tab.~\ref{tab:amp_AAAA}.
\s

As another tree-level process involving the gauge bosons, we can evaluate the amplitudes of the scattering process, $A_L A_L \rightarrow hh$, which includes two longitudinally polarized massive gauge bosons and two Higgs bosons. 
\begin{figure}[ht!]
\centering
\includegraphics[scale=1.0]{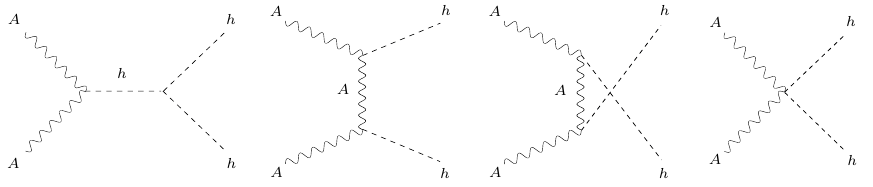}
\caption{Feynman diagrams for the tree-level process $AA\rightarrow hh$}
\label{fig:AAhh_diagrams}
\end{figure}
This process involves the contact-interaction Feynman diagram, as well as those for the $s$-, $t$-, and $u$-channels, as shown in Fig.~\ref{fig:AAhh_diagrams}. In contrast to the process $A_L A_L \rightarrow A_L A_L$, the $t$- and $u$-channel diagrams in the process $A_L A_L \rightarrow hh$ include gauge-boson exchange, seemingly leading to differences in the amplitudes derived from the 5V and 4V $R_\xi$ gauges, even though the sum of all the amplitudes for each diagram is identical regardless of the gauge choice. To check this, let us first compute the following identity,
\begin{align}
&\Gamma^{[AAh]}_{\mu_2 4}(P_2,Q)\,
(-g^{44})\,
\Gamma^{[AAh]}_{4 \mu_1 }(-Q^*,P_1)
\nonumber
\\
&=\Gamma^{[AAh]}_{\mu_2 \sigma}(P_2,Q)\,
\bigg(\frac{Q^{\sigma}Q^{\rho*}}{m^2}\bigg)\,
\Gamma^{[AAh]}_{\rho \mu_1 }(-Q^*,P_1),
\end{align}
in terms of the $t$- or $u$-channel five-momentum $Q=T$ or $U$ defined in Eqs.~\eqref{eq:AAAA_T} and \eqref{eq:AAAA_U}, respectively. These identity enables us to transform the $t$- and $u$-channel amplitudes from the 5V to those in the 4V $R_\xi$ gauge as 
\begin{align}
&\Gamma^{[AAh]}_{\mu_2S}(P_2,Q)\,
\bigg(\frac{-g^{SR}}{Q^2}\bigg)\,
\Gamma^{[AAh]}_{R\mu_1 }(-Q^*,P_1)
\nonumber
\\
&=\Gamma^{[AAh]}_{\mu_2 \sigma}(P_2,Q)\,
\bigg(\frac{-g_{\sigma \rho} +\frac{(1-\xi)Q^\sigma Q^{\rho*}}{Q^2+(1-\xi)m^2}}{Q^2}\bigg)\,
\Gamma^{[AAh]}_{\rho \mu_1 }(-Q^*,P_1)
\nonumber
\\
&\quad +
\Gamma^{[AAh]}_{\mu_2 4}(P_2,Q)\,
\bigg(\frac{1}{Q^2+(1-\xi)m^2}\bigg)\,
\Gamma^{[AAh]}_{4 \mu_1 }(-Q^*,P_1),
\label{eq:equivalence_between_5V_and_4V}
\end{align}
in  terms of the $t$- or $u$-channel five-momentum $Q = T$ or $U$, where $\rho$ and $\sigma$ are the four-vector indices. The identity in Eq.~\eqref{eq:equivalence_between_5V_and_4V} is valid regardless of the gauge-fixing parameter $\xi$ in the 4V $R_\xi$ gauge. This explicitly demonstrates the validity of our approach in treating the Goldstone field as the fifth gauge-field component. We notice that setting the parameter $\xi=1$ leads the 5V $R_\xi$ gauge to become the Feynman gauge, as it does in the 4V $R_\xi$ gauge, suggesting that the renormalizability of the 4V $R_\xi$ gauge~\cite{Faddeev:1967fc} extends to the 5V case.
\s

Next, we find the identity allowing the transformation of the $t$- and $u$-channel amplitudes from the 5V $R_\xi$ gauge to those from the recently developed FD gauge as follows:
\begin{align}
&\Gamma^{[AAh]}_{\mu_2S}(P_2,Q)\,
(-g^{SR} )\,
\Gamma^{[AAh]}_{R\mu_1 }(-Q^*,P_1)
\nonumber
\\
&=\Gamma^{[AAh]}_{\mu_2S}(P_2,Q)\,
\bigg(-g^{SR}+\frac{\tilde{\epsilon}^S(q,0)Q^{R*}+
Q^{S}\tilde{\epsilon}^{R*}(q,0)}{|\sqrt{Q^\mu Q^*_{\mu}}|}\bigg)\,
\Gamma^{[AAh]}_{R\mu_1 }(-Q^*,P_1),
\label{eq:equivalence_5V_FD}
\end{align}
valid due to the divergence-free vertices in Eqs.~\eqref{eq:divergence_free_incoming} and \eqref{eq:divergence_free_outgoing}, where $Q^\mu Q^*_{\mu}$ $(\mu=0,1,2,3)$ is the inner product for the four-vector indices of the five-momentum $Q=T$ or $U$. 
The term in parenthesis at the second line is the numerator of the propagator defined in the FD gauge~\cite{Hagiwara:2020tbx,Chen:2022gxv,Chen:2022xlg,Hagiwara:2024xdh,Hagiwara:2024ron}. Here, $\tilde{\epsilon}(0)$ is given by
\begin{align}
\tilde{\epsilon}^M(p,0)
=
\frac{1}{m}\big(|\vec p\,|, p^0\hat{p},0\big)
-\frac{1}{|\sqrt{p^2}|}\big(p^0,\vec p,0\big),
\end{align}
in terms of a four-momentum $p^\mu$ of a gauge boson.

\section{Identification of the high-energy divergence in tree-level amplitudes}
\label{sec:identification_high-energy}

Using the interaction vertices from Eqs.~\eqref{eq:polar_vertex1} to \eqref{eq:polar_vertex4} and the computational details provided in Appendix.~\ref{appendix:kinematics}, we can directly calculate the amplitudes for the processes $A_LA_L \rightarrow A_LA_L$ and $A_LA_L \rightarrow hh$, as listed in the second columns of Tabs.~\ref{tab:amp_AAAA} and \ref{tab:amp_AAhh}, respectively.
\begin{table}[htb!]
\vskip 0.1cm
\centering
\begin{tabular}{|c|c|c|}
\hline
&&
\\[-10pt]
\; $A_LA_L\rightarrow A_L A_L$ \;
&  \;Polar basis: $\frac{1}{\sqrt{2}}He^{i A^4/v}$\;
&\; Cartesian basis: $\frac{1}{\sqrt{2}}
(\boldsymbol{H}+i\boldsymbol{A}^{4})$\;
\\[5pt] 
\hline
&&
\\[-10pt]
$s$ channel  
& 
$\displaystyle-\frac{\lambda (s\beta^2+2m^2)^2}{2m_h^2(s-m_h^2)}$
&
$\displaystyle-\frac{\lambda(\beta^2m_h^2+2m^2)^2}{2m_h^2(s-m_h^2)}$
%
%
\\[12pt] 
\hline
&&
\\[-10pt]
$t$ channel  
&
$\displaystyle-\frac{\lambda(t+2m^2\cos\theta)^2}{2m_h^2(t-m_h^2)}$
&
$\displaystyle-\frac{\lambda[2m^2(2t+s \cos\theta)+s\beta^2 m_h^2]^2}{s^2 m_h^2(t-m_h^2)}$
%
%
\\[12pt]
\hline
&&
\\[-10pt]
$u$ channel  
&
$\displaystyle-\frac{\lambda (u-2m^2\cos\theta)^2}{2m_h^2(u-m_h^2)}$
&
$\displaystyle-\frac{\lambda[2m^2(2u-s \cos\theta)+s\beta^2 m_h^2]^2}{s^2 m_h^2(u-m_h^2)}$
%
%
\\[12pt]
\hline
&&
\\[-10pt]
contact  
&
-
&
$\displaystyle-\frac{\lambda(3 s\beta^2 m_h^2 +16 m^4)}{2s m^2}$
%
%
\\[12pt]
\hline
&&
\\[-10pt]
\makecell{
sum
\\
in the limit 
\\
$s\rightarrow \infty$
}
&
$-\frac{3}{2}\lambda$
&
$-\frac{3}{2}\lambda$
%
%
\\[18pt]
\hline
\end{tabular}
\caption{Analytic structures of the tree-level amplitudes for the process $A_LA_L \rightarrow A_LA_L$ in the polar (Right) and Cartesian (Left) bases, computed in the c.m. frame. $\beta=\sqrt{1-4m^2/s}$ is the speed of a massive gauge boson in the c.m. frame. The definitions of $t$ and $u$ are given in Eqs.~\eqref{eq:AAAA_t_mandel} and \eqref{eq:AAAA_u_mandel}.}
\label{tab:amp_AAAA}
\end{table}
In the high-energy limit, $\sqrt{s} \rightarrow \infty$, the amplitudes for each diagram diverge. However, their total sum remains finite, which is a well-known phenomenon referred to as the gauge cancellation~\cite{Lee:1977eg}. In this section, we identify the origin of these divergences, ensuring no gauge cancellation in the Cartesian basis for the complex Higgs field. We then show that this framework provides the rationale for the divergences in the 4V $R_\xi$ gauge and clarifies the origin of the absence of gauge cancellation in the FD gauge.
\s

Recall that the masses, $m=ev$ and $m_h=\lambda^{1/2}v/\sqrt{2}$ of the gauge and Higgs bosons are defined to be proportional to the VEV $v$ in Eq.\eqref{eq:5V_Lagrangian}. It is straightforward to verify that the vanishing VEV limit ($v\rightarrow 0$) of each amplitude in Tabs.\ref{tab:amp_AAAA} and \ref{tab:amp_AAhh} is equivalent to its high-energy limit $(\sqrt{s}\rightarrow \infty)$, resulting in the divergence of each amplitude. This is natural because any finite energy scale is always high compared to the vanishing VEV value. In the limit $(v\rightarrow 0)$, the Lagrangian in Eq.~\eqref{eq:5V_Lagrangian} is reduced as
\begin{align}
\mathcal{L}(x)=&-\frac{1}{4} F^{MN}F_{MN}+\frac{1}{2} (\partial^{\mu} h)^2 
\nonumber
\\
&+\bigg(e^2 \epsilon \,h+\frac{e^2}{2}h^2\bigg) \bigg(A^{\mu}-\frac{1}{e\epsilon}\partial^\mu A^4 \bigg)^2 - \frac{\lambda}{4} \epsilon\, h^3 - \frac{\lambda}{16} h^4,
\label{eq:5V_Lagrangian_VEVless}
\end{align}
in terms of the infinitesimal VEV $v=\epsilon$. Given that the interaction vertices are constructed from the Lagrangian, we can conclude that the divergence of each amplitude originates from the diverging interaction strength between the fifth gauge-field component and the Higgs field. We observe that the vanishing strength for the Higgs cubic interaction results in the finite $s$-channel amplitude for the process $A_LA_L\rightarrow hh$.
\s

However, despite the divergence of each amplitude in the vanishing VEV limit, the sum of each amplitude remains finite, as in the case of the high-energy limit. Furthermore, even in the vanishing coupling limit ($e \rightarrow 0$) with the finite VEV, the sum of each amplitude remains still finite, even though this limit implies the full decoupling between the gauge and Higgs bosons. Now, we are at the tree-level processes, so we can find the origin of the finite sum from the perspective on the classical field theory.
\begin{table}[htb!]
\vskip 0.1cm
\centering
\begin{tabular}{|c|c|c|}
\hline
&&
\\[-10pt]
\; $A_LA_L\rightarrow h h$ \;
&  \;Polar basis: $\frac{1}{\sqrt{2}}He^{i A^4/v}$\;
&\; Cartesian basis: $\frac{1}{\sqrt{2}}
(\boldsymbol{H}+i\boldsymbol{A}^{4})$\;
\\[5pt] 
\hline
&&
\\[-10pt]
$s$ channel  
& 
$\displaystyle\frac{3\lambda (s\beta^2+2m^2)}{2(s-m_h^2)}$
&
$\displaystyle\frac{3\lambda (2  m^2 + m_h^2 \beta^2)}{2(s-m_h^2)}$
%
%
\\[12pt] 
\hline
&&
\\[-10pt]
$t$ channel  
&
$\displaystyle-\frac{\lambda s (4t+s\beta^4-s \beta_h^2 \cos^2\theta-4m_h^2 )}{8m_h^2(t-m^2)}$
&
\makecell{
$\displaystyle\frac{\lambda}{4m_h^2(t-m^2)}[m^2(3s \beta^2+8m^2)$
\\[10pt]
$\displaystyle+\beta \beta_h \cos\theta (s m^2 \beta^2-8m^2 m_h^2)$
\\[3pt]
$\displaystyle+8 m^4 \beta_h^2  \cos^2\theta+2m_h^2 \beta^2 (m^2+m_h^2)]$
}
%
%
\\[28pt]
\hline
&&
\\[-10pt]
$u$ channel  
&
$\displaystyle-\frac{\lambda s (4u+s\beta^4-s \beta_h^2 \cos^2\theta-4m_h^2 )}{8m_h^2(u-m^2)}$
&
\makecell{
$\displaystyle\frac{\lambda}{4m_h^2(u-m^2)}[m^2(3s \beta^2+8m^2)$
\\[10pt]
$\displaystyle-\beta \beta_h \cos\theta (s m^2 \beta^2-8m^2 m_h^2)$
\\[3pt]
$\displaystyle+8 m^4 \beta_h^2  \cos^2\theta+2m_h^2 \beta^2 (m^2+m_h^2)]$
}
%
%
\\[28pt]
\hline
&&
\\[-10pt]
contact  
&
$\displaystyle\frac{\lambda (s\beta^2+2m^2)}{2m_h^2}$
&
$\displaystyle\frac{\lambda (s\beta^2 m_h^2 +8m^4)}{2s m_h^2}$
%
%
\\[12pt]
\hline
&&
\\[-10pt]
\makecell{sum
\\
in the limit 
\\
$s\rightarrow \infty$
}
&
$\displaystyle \frac{\lambda}{2}\bigg(1-\frac{2m^2(3+\cos^2\theta)}{m_h^2\sin^2\theta}\bigg)$
&
$\displaystyle \frac{\lambda}{2}\bigg(1-\frac{2m^2(3+\cos^2\theta)}{m_h^2\sin^2\theta}\bigg)$
%
\\[18pt]
\hline
\end{tabular}
\caption{Analytic structures of the tree-level amplitudes for the process $A_LA_L \rightarrow A_LA_L$ in the polar (Right) and Cartesian (Left) bases, computed in the c.m. frame. $\beta=\sqrt{1-4m^2/s}$ and $\beta_h=\sqrt{1-4m_h^2/s}$ are the speeds of a massive gauge boson and a Higgs boson in the c.m. frame, respectively. The definitions of $t$ and $u$ are given in Eqs.~\eqref{eq:AAhh_t_mandel} and \eqref{eq:AAhh_u_mandel}.}
\label{tab:amp_AAhh}
\end{table}
The finiteness implies that the classical Lagrangian describing physical phenomena is also finite. Since fields are not quantized in the classical field theory, any field can be redefined regardless of its dimensionality. Therefore, we cast the fifth gauge-field component as
\begin{align}
A^4(x)=ev \alpha(x),
\label{eq:defintion_A4}
\end{align}
in terms of a dimensionless scalar field $ \alpha $ with the coupling $ e $ and VEV $ v $, resulting in the finite Lagrangian in Eq.~\eqref{eq:5V_Lagrangian_VEVless} in the vanishing VEV limit $(v\rightarrow 0)$. 
\s

Recall that the high-energy divergence of each amplitude for the processes $A_LA_L\rightarrow A_LA_L$ and $A_LA_L\rightarrow hh$ is generated by the quantization not containing the proportionality of the fifth component to the coupling $e$ and VEV $v$. To avoid such high-energy divergence, we can redefine the complex Higgs field in the Cartesian basis as
\begin{align}
\Phi = \frac{1}{\sqrt{2}}(\boldsymbol{H}+i\boldsymbol{A}^{4}),
\label{eq:cartesian_basis}
\end{align}
in the Lagrangian~\eqref{eq:Lagrangian}, with the redefined Higgs field $\boldsymbol{H}$ and fifth gauge-field component $\boldsymbol{A}^4$. The Higgs and fifth-component fields in the polar basis $\Phi=\frac{1}{\sqrt{2}}He^{iA^4/v}$ with $H=v+h$ can be expressed by 
the combinations of $\boldsymbol{H}$ and $\boldsymbol{A}^4$ as
\begin{align}
A^4&=v \sin^{-1}\bigg(\frac{\boldsymbol{A}^{4}}
{\boldsymbol{H}^2+(\boldsymbol{A}^4)^2}\bigg)
=v \cos^{-1}\bigg(\frac{\boldsymbol{H}}
{\boldsymbol{H}^2+(\boldsymbol{A}^{4})^2}\bigg),
\label{eq:cartesian_A4}
\\
H&=\sqrt{\boldsymbol{H}^{2}+(\boldsymbol{A}^{4})^2},
\label{eq:cartesian_H}
\end{align}
reflecting the dependence of $A^4$ on the VEV $v$. To investigate whether the gauge field can be quantized with the three transverse polarization vectors in Eqs.~\eqref{eq:conventional_transverse} and \eqref{eq:third_transverse} as in the case of polar basis, we first expand the Lagrangian with the Higgs field $\boldsymbol{H}=v+\boldsymbol{h}$ after SSB as
\begin{align}
\mathcal{L}(x)
=&-\frac{1}{4} (\boldsymbol{F}^{MN})^2+\frac12 (\partial^\mu \boldsymbol{h})^2
-\frac12 m_h^2 \boldsymbol{h}^{2}
\nonumber
\\
&+\frac{\lambda^{1/2}}{\sqrt{2}m_h}\boldsymbol{h}
\bigg(m^2 \boldsymbol{A}^\mu \boldsymbol{A}_\mu - m
(2\boldsymbol{A}^\mu \partial_\mu \boldsymbol{A}^4+\boldsymbol{A}^4\partial_\mu \boldsymbol{A}^\mu)
-\frac{m_h^2}{2} (\boldsymbol{A}^4 )^2
\bigg)
\nonumber
\\
&+\frac{\lambda}{8} \boldsymbol{h}^{2} 
\bigg(\frac{2m^2}{m_h^2} \boldsymbol{A}^\mu \boldsymbol{A}_{\mu}
- (\boldsymbol{A}^4)^2\bigg)
-\frac{\lambda^{1/2}m_h}{2\sqrt{2}} \boldsymbol{h}^{3}-\frac{\lambda}{16} \boldsymbol{h}^{4} 
\nonumber
\\
&+\frac{\lambda}{16} (\boldsymbol{A}^4)^2 
\bigg(\frac{4 m^2}{m_h^2} \boldsymbol{A}^\mu \boldsymbol{A}_{\mu} 
- (\boldsymbol{A}^4)^2\bigg),
\label{eq:Lagrangian_cartesian}
\end{align}
with $\boldsymbol{F}^{MN}=\partial^M \boldsymbol{A}^N-\partial^N \boldsymbol{A}^M$ where the 4V components $\boldsymbol{A}^\mu$ of the gauge field is equivalent to the polar-basis ones $A^\mu$. The BRST invariance of the Lagrangian ($\mbox{\bf s} \mathcal{L}(x)=0$) can be constructed by defining the BRST transformation of each field as
\begin{align}
\mbox{\bf s} \boldsymbol{A}^{\mu} &=[iQ_B,\boldsymbol{A}^\mu]
= \partial^\mu c,
\label{eq:brst_cartesian1}
\\
\mbox{\bf s} \boldsymbol{A}^{4} &=[iQ_B,\boldsymbol{A}^4]
=e(v+\boldsymbol{h})c,
\label{eq:brst_cartesian2}
\\
\mbox{\bf s}\, \boldsymbol{h}&=[iQ_B,\boldsymbol{h}]
=-e \boldsymbol{A}^{4} c,
\label{eq:brst_cartesian3}
\\
\mbox{\bf s} B&=[iQ_B,B]=0,
\label{eq:brst_cartesian4}
\\
\mbox{\bf s}\, \bar{c}&=\{iQ_B,\bar{c}\}=B,
\label{eq:brst_cartesian5}
\\
\mbox{\bf s}\, c&=\{iQ_B,c\}=0,
\label{eq:brst_cartesian6}
\end{align}
in terms of the BRST operator $\mbox{\bf{s}}$ and the BRST charge $Q_B$. Note that the Lagrangian in Eq.~\eqref{eq:Lagrangian_cartesian} is not invariant under the gauge transformation $\boldsymbol{A}^M \rightarrow \boldsymbol{A}^M+\partial^M \boldsymbol{\Lambda}$ for a scalar field $\boldsymbol{\Lambda}$, resulting in the complicated forms of the BRST transformations compared to those in the polar basis in Eqs.~\eqref{eq:brst_polar1} to \eqref{eq:brst_polar5}. Given that the transformations in Eq.~\eqref{eq:brst_cartesian1} to \eqref{eq:brst_cartesian6} encodes the invariance of the Lagrangian in an interaction theory, we can take the limits $e\rightarrow 0$ and $v\rightarrow \infty$, so $m_h \rightarrow \infty$, but holding the mass $m$ to be finite, leading to the free theory with no interaction terms between the gauge and Higgs fields. In this case, the BRST transformations for each field take the identical forms in those in Eqs.~\eqref{eq:brst_polar1} to \eqref{eq:brst_polar5}, enabling us to quantize equivalently to the case of polar basis.
\s

Employing the expanded Lagrangian~\eqref{eq:Lagrangian_cartesian} in the Cartesian basis, we can construct the interaction vertices as
\begin{align}
\Gamma^{[\boldsymbol{A}\boldsymbol{A}\boldsymbol{h}]}_{M_1M_2}
(P_1,P_2)
&=\frac{\lambda^{1/2}}{\sqrt{2}m_h}
\begin{pmatrix}
2m^2 g_{\mu_1\mu_2}
& \displaystyle i m (2P_{2\mu_1}+P_{1\mu_1})
\\
\displaystyle i m (2P_{1\mu_2}+P_{2\mu_2})
&\displaystyle-m_h^2
\end{pmatrix},
\label{eq:cartesian_vertex_AAh}
\\[8pt]
\Gamma^{[\boldsymbol{A}\boldsymbol{A}\boldsymbol{h}\boldsymbol{h}]}_{MN}
&=\frac{\lambda}{2}
\begin{pmatrix}
\displaystyle\frac{2 m^2}{m_h^2}g_{\mu\nu}&0
\\
0&-1
\end{pmatrix},
\end{align}
for the interactions between the gauge and Higgs bosons, and
\begin{align}
\Gamma^{[\boldsymbol{A}\boldsymbol{A}\boldsymbol{A}\boldsymbol{A}]}_{MNRS}&=-\frac{\lambda m^2}{m_h^2}
(g_{+ MN}g_{- RS}+g_{+ MR}g_{-NS}
+g_{+ MS}g_{+ NR}
\nonumber
\\
&\qquad\qquad +
g_{+ NR}g_{- MS}+g_{+ NS}g_{- MR}+g_{+ RS}g_{+ MN})
\nonumber
\\
&\quad -\frac{3\lambda}{2}g_{- MN}g_{- RS},
\\
\Gamma^{[\boldsymbol{h}\boldsymbol{h}\boldsymbol{h}]}
&=-\frac{3\lambda^{1/2} m_h}{\sqrt{2}},
\\
\Gamma^{[\boldsymbol{h}\boldsymbol{h}\boldsymbol{h}\boldsymbol{h}]}
&=-\frac{3}{2}\lambda,
\end{align}
for the self interactions of the gauge and Higgs bosons, respectively, in terms of two arbitrary 5V incoming momenta $P_{1,2}$. Here, we introduce two five-dimensional rank-2 tensors $g_{+MN}=\mbox{diag}(1,-1,-1,-1,0)$ and $g_{-MN}=\mbox{diag}(0,0,0,0,-1)$. Unlike the case of polar basis, we encounter the four-point self-interaction vertex of the gauge boson (see the diagram at the rightmost panel in Fig.~\ref{fig:AAAA_diagrams}), reflecting that the redefined fields, $\boldsymbol{H}$ and $\boldsymbol{A}^4$, are the mixtures of the fields, $H$ and $A^4$, respectively, as shown in Eq.~\eqref{eq:cartesian_A4}. 
\begin{figure}[ht!]
\centering
\includegraphics[scale=0.5]{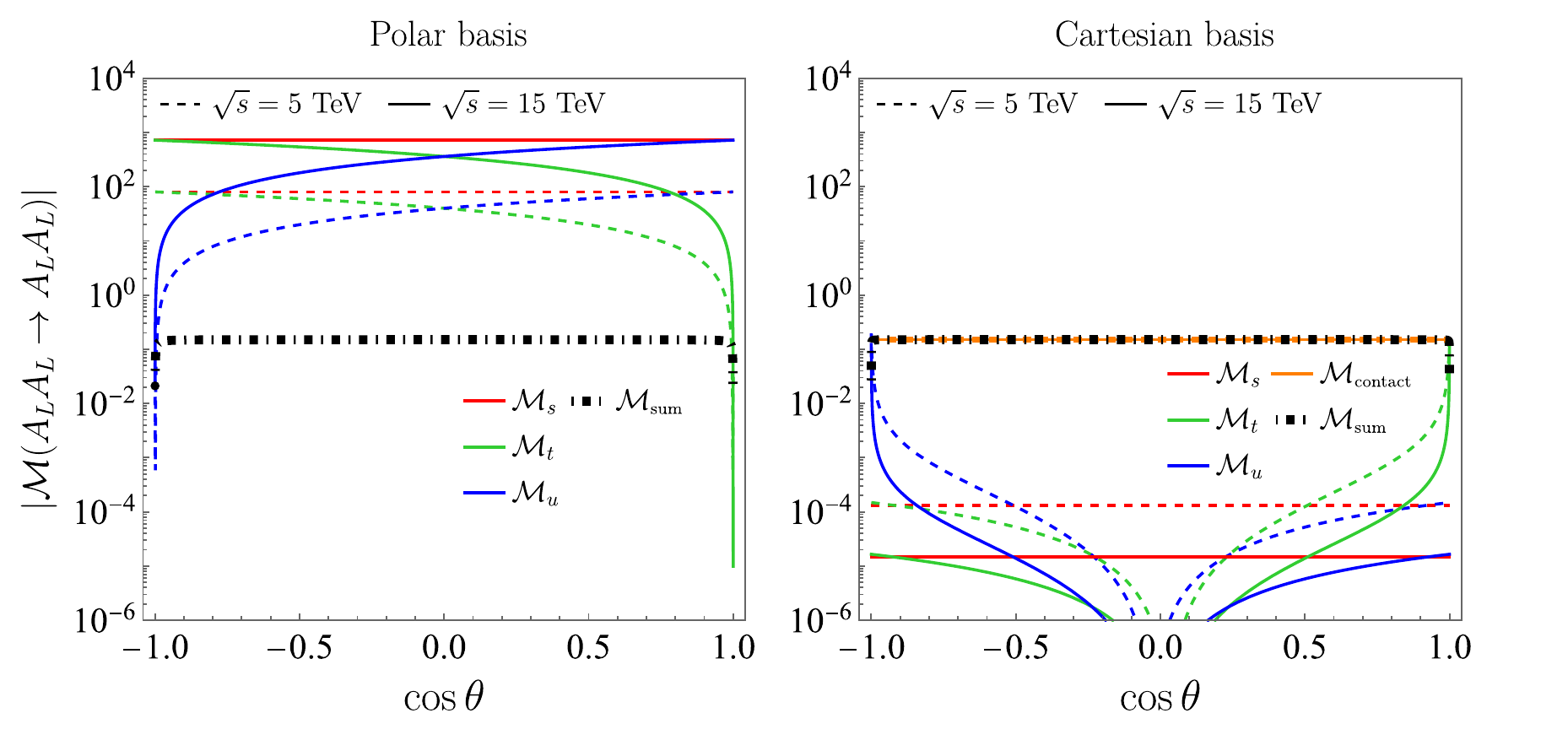}
\caption{Plots for the magnitudes of individual tree-level amplitudes and their sums for the tree-level process $A_LA_L\rightarrow A_LA_L$ in the polar (Left) and the Cartesian (Right) bases.}
\label{fig:polar_cart_AAAA}
\end{figure}
The VEV dependence of the fifth component $A^4$ in the polar basis is reflected in the Cartesian basis, resulting in no divergence of each Lagrangian term. It eventually causes the convergence of each amplitude for the processes $A_LA_L\rightarrow A_LA_L$ and $A_LA_L \rightarrow hh$ in the high-energy limit $\sqrt{s} \rightarrow \infty$, as explicitly shown at the third columns in Tabs.~\ref{tab:amp_AAAA} and \ref{tab:amp_AAhh}, respectively.
\s

\begin{figure}[ht!]
\centering
\includegraphics[scale=0.5]{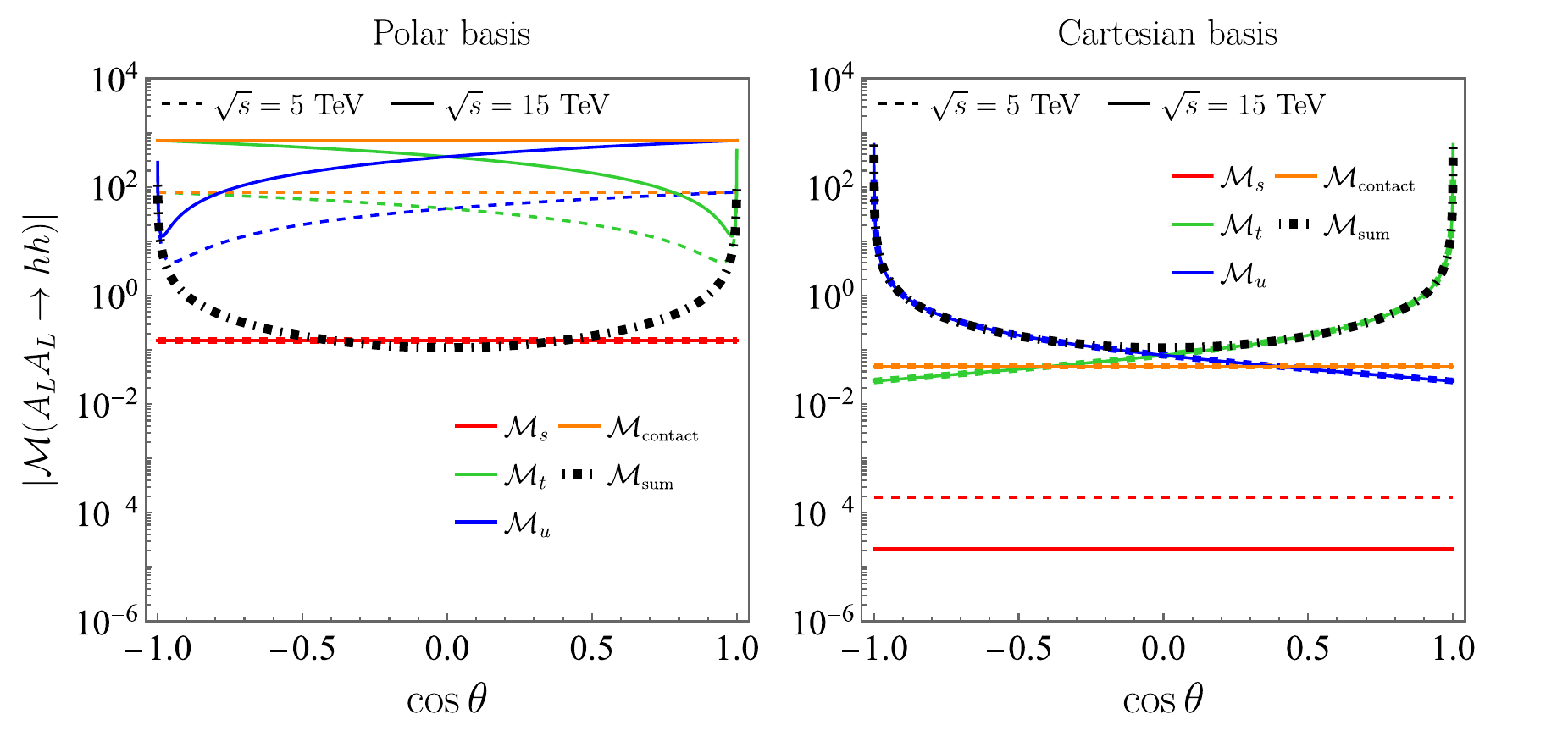}
\caption{Plots for the magnitudes of individual tree-level amplitudes and their sums for the tree-level process $A_LA_L\rightarrow hh$ in the polar (Left) and the Cartesian (Right) bases.}
\label{fig:polar_cart_AAhh}
\end{figure}

To explicitly demonstrate the resolution of high-energy divergences, we present the magnitudes of individual amplitudes and their sums for the $A_LA_L \rightarrow A_LA_L$ and $A_LA_L \rightarrow hh$ processes at center-of-mass energies $\sqrt{s} = 5$ and $15$ TeV in Figs.\ref{fig:polar_cart_AAAA} and \ref{fig:polar_cart_AAhh}, with the parameters $\lambda = 0.1$, $m = 91$ GeV, and $m_h = 125$ GeV.
To simplify the analysis, we refer to the magnitude of each amplitude simply as the amplitude, for this discussion. In each panel, the dashed lines represent the amplitudes at $\sqrt{s} = 5$ TeV, while the solid lines correspond to those at $\sqrt{s} = 15$ TeV. For the $A_LA_L \rightarrow A_LA_L$ process shown in Fig.~\ref{fig:polar_cart_AAAA}, the amplitudes in the polar basis increase with higher collision energy, but their sum, depicted by the black dotted-dashed line, is significantly smaller than each individual amplitude, indicating strong gauge cancellation. In contrast, the amplitudes in the Cartesian basis decrease with higher collision energy, except for the contact-interaction amplitude, which dominates the sum. A similar pattern is observed for the $A_LA_L \rightarrow hh$ process in Fig.\ref{fig:polar_cart_AAhh}. In the polar basis, each amplitude increases with higher collision energy except for the $s$-channel amplitude, but their sum remains much smaller than any increasing amplitudes with higher collision energy. However, in the Cartesian basis, no individual amplitude exceeds their sum, indicating the absence of gauge cancellation in the Cartesian basis.
\s

What the Lagrangian in Eq.~\eqref{eq:Lagrangian_cartesian} is not invariant under the gauge transformation $\boldsymbol{A}^M\rightarrow \boldsymbol{A}^M + \partial^M \boldsymbol{\Lambda}$ results in the loss of the divergence-free nature of each vertex unlike the case of polar basis, as shown in Eqs.~\eqref{eq:divergence_free_incoming} and \eqref{eq:divergence_free_outgoing}. However, we can still demonstrate the transformation of the amplitudes from the 5V $R_\xi$ gauge to those from the 4V $R_\xi$ gauge, because the gauge invariance under the transformation $A^M\rightarrow A^M + \partial^M \Lambda$ in the polar basis is still implicitly encoded in the Lagrangian, ensured by the BRST symmetry. Thus, the sum of amplitudes for each diagram, coupled to the polarization vectors of the external gauge bosons, is divergence-free, resulting in the Ward identity in the 5V description as 
\begin{align}
P^M T_M(A)&=0 \quad \mbox{for an incoming $A$},
\\
P^{M*} T_M(A)&=0 \quad \mbox{for an outgoing $A$},
\end{align}
reflecting the conservation of any current coupled to the gauge field, where $P^M$ and its complex conjugation $P^{M*}$ are the five-momenta of the incoming and outgoing external gauge bosons, respectively, and $T_M(A)$ stands for the full amplitude of a process involving at least one gauge boson $A$, which is coupled to its polarization five-vectors. The independence of physical phenomena from the choice of basis can be demonstrated directly by comparing the sums of the $A_L A_L \rightarrow A_L A_L$ and $A_L A_L \rightarrow hh$ scattering amplitudes in the two bases, yielding identical results, as shown in the last rows of Tabs.~\ref{tab:amp_AAAA} and \ref{tab:amp_AAhh}.
\s

To reach the transformation, let us consider the $t$- and $u$-channel amplitudes including the gauge-boson propagator for the process $A_LA_L \rightarrow hh$, as discussed in Sec.~\ref{sec:equivalence_5V_4V} in the polar basis. Then, we can find the following identities,
\begin{align}
Q^{R*} \Gamma^{[\boldsymbol{A}\boldsymbol{A}\boldsymbol{h}]}_{R M_1}(-Q^*, P_1)
&=\frac{2\sqrt{2}\eta m^2}{m_h} P^*_{1 M_1},
\label{eq:Q_contraction1}
\\
Q^{S} \Gamma^{[\boldsymbol{A}\boldsymbol{A}\boldsymbol{h}]}_{M_2 S}(P_2, Q )  
&=\frac{2\sqrt{2}\eta m^2}{m_h} P^*_{2 M_2},
\label{eq:Q_contraction2}
\end{align}
for the $t$- or $u$-channel five-momentum $Q=T$ or $U$ in Eqs.~\eqref{eq:AAhh_T} and \eqref{eq:AAhh_U}, respectively. The contraction of the vertices with the polarization five-vectors of the external gauge bosons results in the gauge-parameter independence of the $t$- and $u$-channel amplitudes for the process $A_LA_L\rightarrow hh$ in the 5V $R_\xi$ gauge, allowing us to omit the gauge-parameter dependent term $Q^M Q^{N*}/Q^2$ in the 5V $R_\xi$ gauge propagator~\eqref{eq:5V_propagator}. It enables us to find the identity for the propagation of the fifth gauge-field component as
\begin{align}
&\Gamma^{[\boldsymbol{A}\boldsymbol{A}\boldsymbol{h}]}_{\mu_2 4}(P_2,Q)\,
(-g^{44})\,
\Gamma^{[\boldsymbol{A}\boldsymbol{A}\boldsymbol{h}]}_{4 \mu_1 }(-Q^*,P_1)
\nonumber
\\
&=\Gamma^{[\boldsymbol{A}\boldsymbol{A}\boldsymbol{h}]}_{\mu_2 \sigma}(P_2,Q)\,
\bigg(\frac{(Q^{\sigma}+P_2^{\sigma }/2)(Q^{\rho*}-P_1^{\rho *}/2)}{m^2}\bigg)\,
\Gamma^{[\boldsymbol{A}\boldsymbol{A}\boldsymbol{h}]}_{\rho \mu_1 }(-Q^*,P_1),
\end{align}
where $\rho$ and $\sigma$ are the four-vector indices, respectively. Since the vertex $\Gamma^{[\boldsymbol{A}\boldsymbol{A}\boldsymbol{h}]}_{\mu\nu}$  with only the four-vector indices in Eq.~\eqref{eq:cartesian_vertex_AAh} is proportional to the four-dimensional metric tensor, the momenta, $P_1^{\rho *}$ and $P_2^{\sigma}$, are directly contracted to the external polarization vectors. It results in no cancellation of the contracted terms for the third transverse polarization vector~\eqref{eq:third_transverse} due to the contraction only for the four-vector index. The cancellation can be achieved by employing the 5V representation $\epsilon^M_0(0)$ in Eq.~\eqref{eq:5V_rep_of_longitudinal_vector} of the longitudinal mode as it is divergence-free, $P^{M*}\epsilon_{M}(p,0)=0$ and $P^{\mu*}\epsilon_{\mu}(p,0)=0$, for both the five- and four-momentum. Therefore, employing $\epsilon^M_0(0)$ enables us to conclude that the full amplitudes for the processes $A_LA_L\rightarrow A_LA_L$ and $A_LA_L\rightarrow hh$ from the 5V and 4V $R_\xi$ gauges are identical to each other, ensuring the validity of the 5V framework. Additionally, we notice that the equivalence between the full amplitudes from the 5V $R_\xi$ and FD gauges can be checked straightforwardly by employing the identities in Eqs.~\eqref{eq:Q_contraction1} and \eqref{eq:Q_contraction2}, resulting in the cancellation of the additional terms in the FD gauge propagator in Eq.~\eqref{eq:equivalence_5V_FD} except for the five-dimensional metric tensor via the contraction with the polarization five-vectors $\epsilon^{M_{1,2}}(p_{1,2},\lambda_{1,2})$.
\s

\begin{figure}[ht!]
\centering
\includegraphics[scale=0.5]{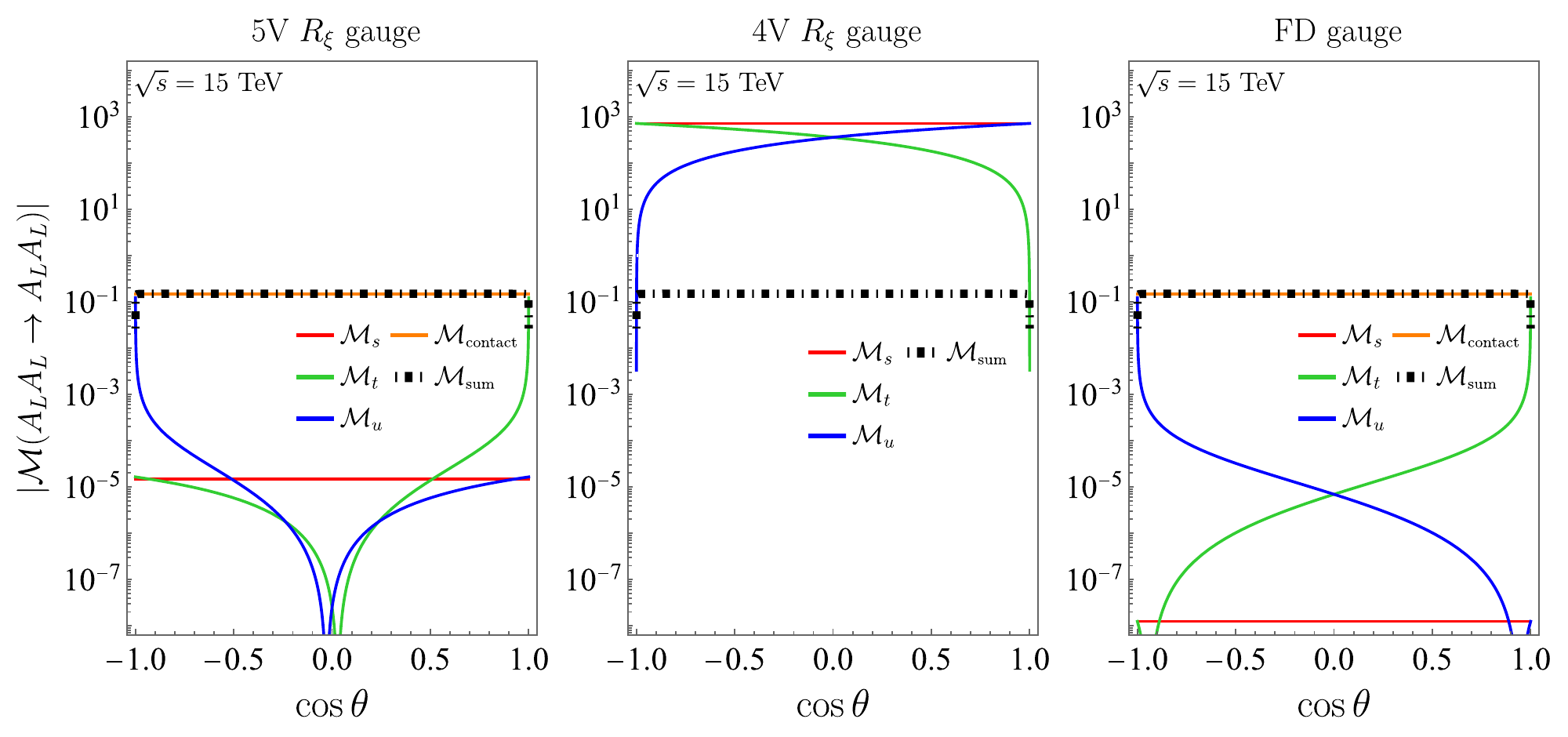}
\caption{Plots for the magnitudes of individual tree-level amplitudes and their sums for the process $A_LA_L \rightarrow A_LA_L$ from the 5V $R_\xi$ (Left), 4V $R_\xi$ (Middle), and FD (Right) gauges in the Cartesian basis.}
\label{fig:gauges_AAAA}
\end{figure}

\begin{figure}[ht!]
\centering
\includegraphics[scale=0.5]{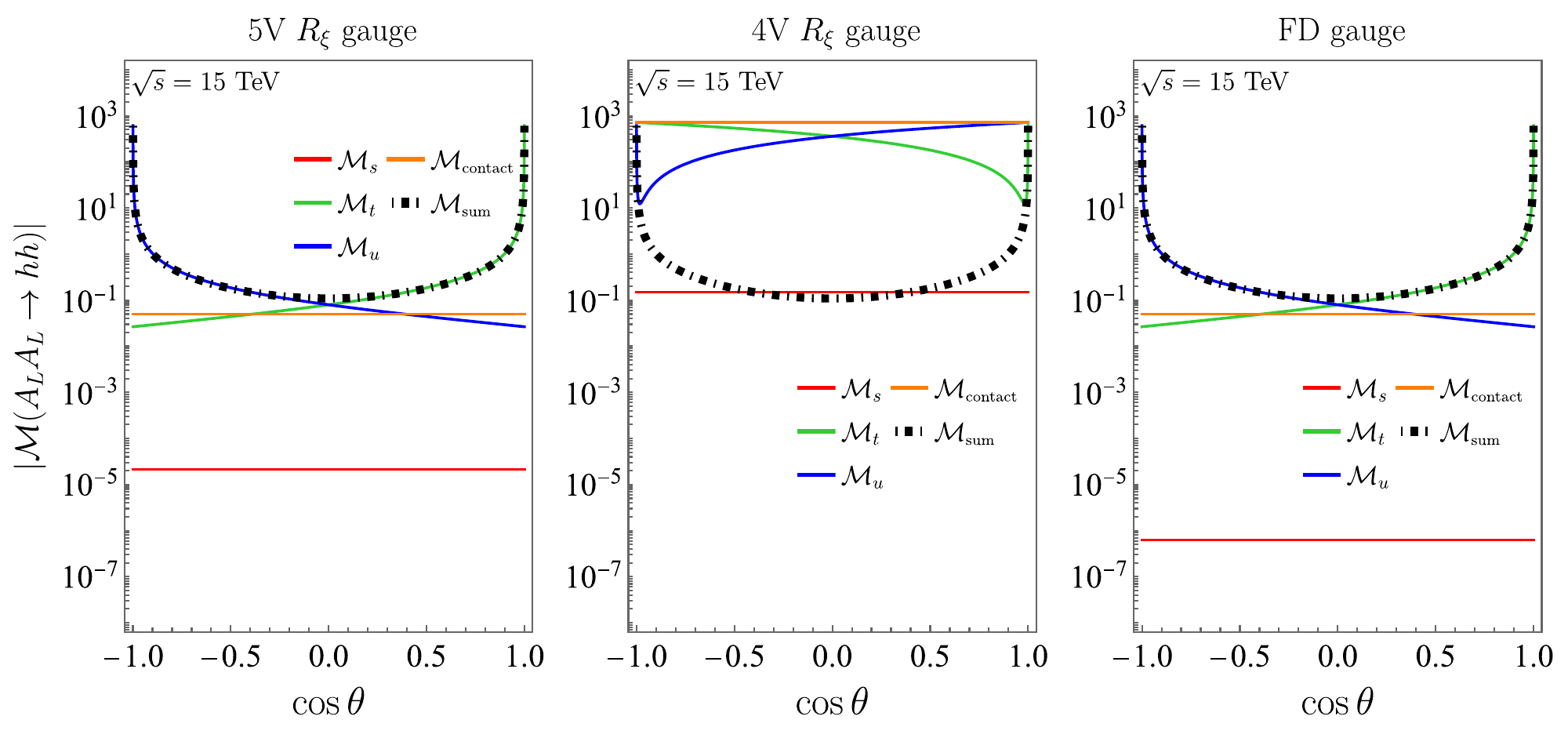}
\caption{Plots for the magnitudes of individual tree-level amplitudes and their sums for the process $A_LA_L \rightarrow hh$ from the 5V $R_\xi$ (Left), 4V $R_\xi$ (Middle), and FD (Right) gauges in the Cartesian basis.}
\label{fig:gauges_AAhh}
\end{figure}

It is worthwhile to explicitly compare the amplitudes from the 5V $R_\xi$, 4V $R_\xi$, and FD gauges to illustrate the impact of the 5V framework. Thus, we include plots for the magnitudes of individual amplitudes and their sums from these gauges for the processes $A_LA_L\rightarrow A_LA_L$ and $A_LA_L \rightarrow hh$ at the center-of-mass collision energy $\sqrt{s}=15$ GeV with the parameters $\lambda=0.1$, $m=91$ GeV, and $m_h=125$ GeV in Figs.\ref{fig:gauges_AAAA} and\ref{fig:gauges_AAhh}. 
Before conducting a detailed analysis of the plots, we first identify the origin of the high-energy divergences in the 4V $R_\xi$ gauge and investigate the absence of gauge cancellation in the FD gauge. In the 5V $R_\xi$, 4V $R_\xi$, and FD gauges, the helicity-0 polarization vectors are written as
\begin{align}
\epsilon^M(p,0)&=\frac{1}{E}(0,0,0,m,-i p),
\label{eq:5V_polarization}
\\
\epsilon^M_{0}(p,0)&
=\epsilon^M(p,0)+\frac{p}{m}\frac{P^M}{E}
=\frac{1}{m}
(p,0,0,E,0),
\label{eq:4V_polarization}
\\
\epsilon^M_{\scriptsize\mbox{FD}}(p,0)&
=\epsilon^M(p,0)+\frac{(p-E)}{m}\frac{P^M}{E}
=\frac{1}{m}
(p-E,0,0,E-p,-im),
\label{eq:FD_polarization}
\end{align}
respectively, for a gauge boson moving along the $z$ axis, with the on-shell energy $E=\sqrt{p^2+m^2}$. For the 4V $R_\xi$ gauge, the polarization vector $\epsilon^M_0(0)$, with an additional five-momentum factor added to $\epsilon^M(0)$, causes high-energy divergences in each amplitude because the interaction vertices in the Cartesian basis are not divergence-free, resulting in the unphysical five-momentum contribution to each amplitude. For the FD gauge, although there is an additional five-momentum factor, it vanishes in the high-energy limit, resulting in no gauge cancellation, similar to the 5V $R_\xi$ gauge.
\s

For simplicity in analyzing the plots in Figs.\ref{fig:gauges_AAAA} and\ref{fig:gauges_AAhh}, we refer to the magnitude of each amplitude simply as the amplitude for this discussion.
For the process $A_LA_L\rightarrow A_LA_L$, only the 4V $R_\xi$ gauge exhibits amplitudes significantly larger than their sum. The amplitudes from the FD gauge show slightly different behavior compared to those of the 5V $R_\xi$ gauge, which originates from the additional five-momentum factor in Eq.~\eqref{eq:FD_polarization}. Similarly, for the $A_LA_L\rightarrow hh$ process, only the 4V $R_\xi$ gauge exhibits amplitudes significantly larger than their sum, and the FD gauge amplitudes deviate slightly from those of the 5V $R_\xi$ gauge, but neither the 5V $R_\xi$ gauge nor the FD gauge results in gauge cancellation.
\s

The definition~\eqref{eq:defintion_A4} of the fifth gauge-field component ensures that in the vanishing VEV ($v\rightarrow 0$) limit, the gauge field is required to be quantized without the third transverse polarization mode due to the proportionality of the fifth component to the VEV. Similarly, in the vanishing coupling limit $e\rightarrow 0$, the third transverse polarization mode is also absent in the quantized gauge field, due to the vanishing of the fifth component, where the Higgs field is a purely real scalar, implying the full decoupling between the gauge and Higgs bosons. Thus, one can observe that the 5V $R_\xi$ gauge-fixing term in Eq.~\eqref{eq:5V_gauge-fixing} is automatically reduced to the 4V $R_\xi$ gauge-fixing term in the massless limit, ($v\rightarrow 0$) or ($e\rightarrow 0$).

\section{Conclusion}
\label{sec:conclusion}

We presented a novel $R_\xi$ gauge in the 5V description within the Abelian Higgs model. By treating the Goldstone field as the fifth gauge-field component, we derived the 5V $R_\xi$ gauge-fixing term by imposing BRST invariance on the Lagrangian. We observed that for the gauge parameter $\xi=1$, the 5V $R_\xi$ gauge is equivalent to the Feynman gauge in the covariant $R_\xi$ gauge within the 4V framework. It strongly suggests the validity of renormalizability of the 5V $R_\xi$ gauge, but should be investigated rigorously for the renormalizability of other $\xi$ values. We demonstrated that in the 5V $R_\xi$ gauge, all physical polarization vectors of a massive gauge boson are transverse to the five-momentum. The structure of the third transverse mode, corresponding to the longitudinal mode in the 4V description, clearly indicates that it is not an intrinsic property of the gauge boson, as its spin-angular momentum transitions from spin-1 in the rest frame to spin-0 in the high-energy or massless limit.
\s

To highlight the utility of the 5V framework, we analyzed the tree-level amplitudes for the processes $A_LA_L \rightarrow A_LA_L$ and $A_LA_L \rightarrow hh$, involving massive gauge bosons with longitudinal modes and Higgs bosons. We showed that in the polar basis for the complex Higgs field, the amplitudes for each Feynman diagram, derived from the 5V $R_\xi$ gauge, are equivalent to those from the 4V $R_\xi$ and FD gauges. In the polar basis, the amplitudes for each diagram diverge in the high-energy limit regardless of the gauge choice. We identified
the origin of the divergences from the equivalence between the high-energy and vanishing VEV limits. The divergences were resolved by adopting the Cartesian basis, which reflects the vanishing of the fifth component in the polar basis as the VEV vanishes. The resolution was quantitatively demonstrated by analyzing the magnitudes of individual amplitudes and their sums, highlighting the absence of gauge cancellation in the Cartesian basis.
\s

Next, we showed that the 5V framework can straightforwardly trace the origin of the high-energy divergences in the amplitudes from the 4V $R_\xi$ gauge, which arises from the unphysical five-momentum contribution. While the FD gauge also includes a five-momentum contribution, it vanishes in the high-energy limit, resulting in no gauge cancellation, similar to the 5V $R_\xi$ gauge. These findings were explicitly demonstrated through a quantitative investigation of the amplitudes and their sums.
\s

The structure of the third transverse polarization vector of the gauge boson is expected to deepen our understanding of the dynamics of massive gauge bosons. The simple analytic structure of the 5V framework, along with the absence of gauge cancellation in the 5V $R_\xi$ gauge, provides a useful criterion for assessing the degree of high-energy divergences in tree-level amplitudes computed in various gauges. Thus, this approach offers significant practical advantages for both analytical and numerical amplitude calculations. Notably, this work presents the first extension in which the asymptotic on-shell gauge boson state is defined together with a Goldstone state multiplied by a non-constant kinematic factor~\cite{Wulzer:2013mza,Cuomo:2019siu}. This observation suggests that a natural next step is to generalize the gauge boson state with an arbitrary kinematic factor at the Goldstone state. Furthermore, this framework can be extended to non-Abelian gauge theories, which incorporate the self-interactions of gauge bosons, thereby paving the way for even broader applications. This extension is currently under investigation and will be reported separately, with potential implications for a wider class of gauge theories.

\section*{Acknowledgments}
We thank Prof. Andrea Wulzer, Prof. Alex Pomarol, Dr. Kiyoharu Kawana, and Dr. Yu-hui Zheng for their valuable comments on this work. This work was supported by a KIAS Individual Grant (QP090001) through the Quantum Universe Center at the Korea Institute for Advanced Study. We also acknowledge the Institut de Física d'Altes Energies (IFAE) for their hospitality during our visit.

\appendix
\renewcommand{\theequation}{A.\arabic{equation}}

\section{Kinematics and Feynman rules in the c.m. frame}
\label{appendix:kinematics}
\setcounter{equation}{0}

This appendix presents the computational ingredients for the tree-level amplitudes of the processes $A_LA_L \rightarrow A_LA_L$ and $A_LA_L \rightarrow hh$ in the c.m. frame, with the kinematic configurations shown in Fig.~\ref{fig:kinematics}. For the process $A_LA_L \rightarrow A_LA_L$, the momenta of the four massive gauge bosons are given by
\begin{align}
P_1 &= \frac{\sqrt{s}}{2}(1, 0, 0, \beta, i m), \quad 
P_2 = \frac{\sqrt{s}}{2}(1, 0, 0, -\beta, i m), \nonumber 
\\
P_3 &= \frac{\sqrt{s}}{2}(1, \beta \sin\theta, 0, \beta \cos\theta, i m), \quad 
P_4 = \frac{\sqrt{s}}{2}(1, -\beta \sin\theta, 0, -\beta \cos\theta, i m),
\end{align}
where the speed of a gauge boson is given by $\beta = \sqrt{1 - 4m^2/s}$ in the c.m. frame. The process $A_LA_L \rightarrow hh$ includes four diagrams: the $s$-, $t$-, and $u$-channels, where the Higgs boson propagates in all three channels; the contact term as shown in Fig.~\ref{fig:AAAA_diagrams}. The propagator momentum of the Higgs boson in each channel can be written in the 5V description as
\begin{align}
S^M &= \big((P_1 + P_2)^\mu, im_h\big) = (\sqrt{s}, 0, 0, 0, im_h),
\label{eq:AAAA_S}
\\
T^M &= \big((P_1 - P_3)^\mu, im_h\big) = \frac{\sqrt{s}}{2}
\big(0, -\beta \sin\theta, 0, \beta(1 - \cos\theta), im_h\big), 
\label{eq:AAAA_T}
\\
U^M &= \big((P_1 - P_4)^\mu, im_h\big) = \frac{\sqrt{s}}{2}
\big(0, \beta \sin\theta, 0, \beta(1 + \cos\theta), im_h\big),
\label{eq:AAAA_U}
\end{align}
where the 5V inner product of each momentum satisfies $S^2 = s - m_h^2$, $T^2 = t - m_h^2$, and $U^2 = u - m_h^2$ with the Mandelstam variables $s$, $t$, and $u$. Here, $t$ and $u$ can be computed as
\begin{align}
t &= -\frac{s\beta^2}{2}(1 - \cos\theta), 
\label{eq:AAAA_t_mandel}
\\
u &= -\frac{s\beta^2}{2}(1 + \cos\theta),
\label{eq:AAAA_u_mandel}
\end{align}
with the simple forms $t \simeq -s(1 - \cos\theta)/2$ and $u \simeq -s(1 + \cos\theta)/2$ in the high-energy limit ($\sqrt{s} \rightarrow \infty$).
\s

For the process $A_LA_L \rightarrow hh$, the five-momenta of the gauge and Higgs bosons are
\begin{align}
P_1 &= \frac{\sqrt{s}}{2}(1, 0, 0, \beta, i m), \quad 
P_2 = \frac{\sqrt{s}}{2}(1, 0, 0, -\beta, i m), \nonumber 
\\
P_3 &= \frac{\sqrt{s}}{2}(1, \beta_h \sin\theta, 0, \beta_h \cos\theta, i m_h), \quad 
P_4 = \frac{\sqrt{s}}{2}(1, -\beta_h \sin\theta, 0, -\beta_h \cos\theta, i m_h),
\end{align}
with the speed $\beta_h = \sqrt{1 - 4m_h^2/s}$ of the Higgs boson in the c.m. frame. The process $A_LA_L \rightarrow hh$ includes three diagrams: the $s$-channel, where the Higgs boson propagates; and the $t$- and $u$-channels, where the gauge bosons propagate; and the contact term as shown in Fig.~\ref{fig:AAhh_diagrams}. The $s$-channel propagator momentum of the Higgs boson and the $t$- and $u$-channel propagator momenta of the gauge bosons are written as
\begin{align}
S^M &= \big((P_1 + P_2)^\mu, im_h\big) = (\sqrt{s}, 0, 0, 0, im_h), 
\label{eq:AAhh_S}
\\
T^M &= \big((P_1 - P_3)^\mu, im\big) = \frac{\sqrt{s}}{2}
\big(0, -\beta_h \sin\theta, 0, \beta - \beta_h \cos\theta, im\big), 
\label{eq:AAhh_T}
\\
U^M &= \big((P_1 - P_4)^\mu, im\big) = \frac{\sqrt{s}}{2}\big(0, \beta_h \sin\theta, 0, \beta + \beta_h \cos\theta, im\big),
\label{eq:AAhh_U}
\end{align}
with the 5V inner product of each five-momentum, $S^2 = s - m_h^2$, $T^2 = t - m^2$, and $U^2 = u - m^2$. Here, the Mandelstam variables $t$ and $u$ are given by
\begin{align}
t &= -\frac12(s - 2m^2 - 2m_h^2 - s\beta\beta_h \cos\theta), 
\label{eq:AAhh_t_mandel}
\\
u &= -\frac12(s - 2m^2 - 2m_h^2 + s\beta\beta_h \cos\theta),
\label{eq:AAhh_u_mandel}
\end{align}
with the simple forms $t \simeq -s(1 - \cos\theta)/2$ and $u \simeq -s(1 + \cos\theta)/2$ in the high-energy limit ($\sqrt{s} \rightarrow \infty$).
\s

In the computations, we employ the following polarization vectors of the incoming gauge bosons,
\begin{align}
\epsilon^M(p_1, \pm 1) &= \frac{1}{\sqrt{2}}(0, \mp 1, -i, 0, 0), \quad 
\epsilon^M(p_1, 0) = \frac{1}{\sqrt{s}}(0, 0, 0, 2m, -i\sqrt{s}\beta), \\
\epsilon^M(p_2, \pm 1) &= \frac{1}{\sqrt{2}}(0, \mp 1, i, 0, 0), \quad 
\epsilon^M(p_2, 0) = \frac{1}{\sqrt{s}}(0, 0, 0, -2m, -i\sqrt{s}\beta),
\end{align}
while those of the outgoing gauge bosons are:
\begin{align}
\epsilon^M(p_3, \pm 1) &= \frac{1}{\sqrt{2}}(0, \mp \cos\theta, -i, \pm \sin\theta, 0), \\
\epsilon^M(p_3, 0) &= \frac{1}{\sqrt{s}}(0, 2m \sin\theta, 0, 2m \cos\theta, -i\sqrt{s}\beta), \\
\epsilon^M(p_4, \pm 1) &= \frac{1}{\sqrt{2}}(0, \mp \cos\theta, i, \pm \sin\theta, 0), \\
\epsilon^M(p_4, 0) &= \frac{1}{\sqrt{s}}(0, -2m \sin\theta, 0, -2m \cos\theta, -i\sqrt{s}\beta).
\end{align}

\bibliographystyle{unsrt}
\bibliography{ref}

\end{document}